# Mobility Gaps between Low-Income and Not Low-Income Households: A Case Study in New York State

Yuandong Liu, Ph.D.,[1] and Majbah Uddin, Ph.D., EIT [2]

[1]National Transportation Research Center, Oak Ridge National Laboratory, 1 Bethel Valley Rd, Oak Ridge, TN 37830; e-mail: liuy@ornl.gov
[2]National Transportation Research Center, Oak Ridge National Laboratory, 1 Bethel Valley Rd, Oak Ridge, TN 37830; e-mail: uddinm@ornl.gov

## ABSTRACT

Understanding the travel challenges faced by low-income residents has always been and continues to be one of the most important transportation equity topics. This study aims to explore the mobility gaps between low-income households (HHs) and not low-income HHs, and how the gaps vary within different socio-demographic population groups in New York State (NYS). The latest National Household Travel Survey data was used as the primary data source for the analysis. The study first employed the K-prototype clustering algorithm to categorize the HHs in NYS based on their socio-demographic attributes. Five population groups were identified based on nine different household (HH) features such as HH size, vehicle ownership, and elderly status of its members. Then, the mobility differences, measured by trip frequency, trip distance, travel time, and person miles traveled were examined among the five population groups. Results suggest that the individuals in low-income HHs consistently took fewer trips and made shorter trips compared to their not low-income counterparts in NYS. The travel distance gaps were most obvious among white HHs with more vehicles than drivers. In addition, while the population from low-income HHs made shorter trips on average (2.7 miles shorter per trip), they experienced longer travel time than those from not low-income HHs (1.8 minutes longer per trip). These key findings provide a deeper understanding of the travel behavior disparities between low-income and not low-income households. The findings could also support policymakers and transportation planners in addressing the critical needs of residents in low-income households in NYS and provide inputs for designing a more equitable transportation system.

## INTRODUCTION

The 2019 Consumer Expenditure Survey shows that transportation cost is the second highest among all expense categories—lower than housing expenditures but higher than food costs (BLS 2019). According to the survey, around 17% of expenditures of a consumer unit were spent on transportation in 2019. This relatively high share of expenditures on transportation poses financial burdens, especially to those in low-income households (HHs). Based on the 2017 National Household Travel Survey (NHTS), lower-income households were more likely to express being cost conscious regarding travel (FHWA 2019a). Constrained by their limited budgets and resources, low-income HHs face more travel challenges than their not low-income counterparts. In general, travels by the low-income HHs are fewer, take more time, traverse less distance, and rely more on alternative modes than higher-income HHs (Clifton 2003; Banerjee 2018). A recent study explored the full array of travel behaviors of the HHs below and above the poverty level in the US (Banerjee 2018). Travel behavior disparities were identified in terms of daily trip rate, trip distance, as well as mode choices. Banerjee showed that on average, the daily person trip rate for low-income HHs was 2.9, in contrast to 3.5 for those in not low-income HHs. Low-income HHs were also likely to travel shorter distances compared to their not low-income counterparts. In addition to daily trips and trip distance, the mode choice decision of low-income individuals was found to be impacted by their limited budgets. In making mode-choice decisions, low-income travelers carefully evaluate the costs of travel (time and out-of-pocket expenses) against the benefits of each mode available to them (Agrawal et al. 2011). Overall, compared to higher income populations, low-income populations are more likely to use public transit and non-motorized transportation modes instead of driving to meet their daily travel needs. Despite this fact, encouraging people to use public transport instead of cars was found to be challenging, no less for low-income HHs than other income HHs (Taylor 2009). A recent report based on 2017 NHTS found that while people's attitudes about the cost of travel were related to their income, people in lower-income households did not agree that they walked or took transit to save money (FHWA 2019b).

Several factors were identified to have impacts on low-income household travel behavior and mobility patterns. Among them, vehicle ownership or having access to privately owned vehicles was one of the most influential factors. Car usage performs a critical role in facilitating access to, and participation in, a wide range of key services and opportunities for low-income HHs (Blumenberg and Pierce 2012; Rozynek et al. 2022; Taylor 2009). While automobile ownership increases personal miles traveled (PMT) for all adults, it is particularly influential in increasing the travel of low-income adults. HHs accrue greater marginal benefits by moving from zero to one vehicle than by purchasing additional vehicles when they already own one or more (Blumenberg and Pierce 2012).

The previous studies have explored the travel behaviors of low-income HHs from diverse perspectives, including differences between different income groups, low-income individuals' travel concerns, their travel decision-making process, and the influencing factors. However, one limitation of the previous studies is that they either treat the low-income HHs as a single group or focused on one or two dimensions of the low-income HHs, such as those who own automobile vehicles vs those without automobile vehicles. Very few study studies were conducted to examine low-income household travel behavior within various demographic groups. To facilitate the understanding of mobility gaps within different population classes, this study conducts a case study in NYS. The study first adopts a K-prototype algorithm-based method to categorize the NYS low-income HHs into representative groups. Then the mobility gaps between low-income and not low-income HHs in NYS are evaluated among these groups based on several mobility measurements.

**DATA SOURCE**

**NHTS**

NHTS (FHWA 2017) is a national travel survey of U.S. households sponsored by the Federal Highway Administration. The survey collects daily travel information that is linked to individual personal and household characteristics, and vehicle attributes, such as trip frequency, travel distance and time, mode of transportation, and trip purpose. The latest NHTS was done in 2017, which surveyed over 129,000 HHs. Among these, 26,000 HHs were from a national sample and the rest were from Add-on samples purchased by thirteen State or MPO partners. The New York State Department of Transportation (NYSDOT) was an Add-on partner in the 2017 NHTS. Consequently, NYSDOT received travel data for over 17,000 HHs in the State. This study was conducted with all the data samples in NYS, including those from the NYS add-on programs.

**Defining Low-Income Households**

A widely adopted approach in the literature to define low-income HHs is to use a simple cutoff value for household income (Lou et al. 2020; Moniruzzaman et al. 2015). Any HHs below the cutoff value (household income threshold) are defined as low-income HHs. This method, while straightforward to implement, does not consider HH properties such as size/composition and location (urban or rural). In this study, five different low-income definitions (other than the simple cutoff value method) used in literature or published by different agencies were explored and examined, as presented in Table 1. Generally, these thresholds are updated on a yearly basis. To be consistent with the 2017 NHTS, the low-income definitions for 2017 were compared. Comparing the six different low-income household thresholds, the first two thresholds—Census poverty threshold and Health and Human Services (HHS) poverty guidelines are established at the national level without considering the cost of living and housing market in different areas (e.g., urban and rural areas). As presented in the table, their income thresholds for 4-person low-income HHs in New York City (NYC) are the lowest among all definitions. The other three thresholds, Census Bureau Supplementary Poverty Measure (SPM), Lower Living Standard Income Level (LLSIL) Guidelines and Housing and Urban Development (HUD) income limit consider the regional differences and establish the low-income threshold in finer resolution level such as metropolitan areas. Among these three definitions, the low-income threshold developed by HUD provides the highest geographical resolution (at county level or metropolitan area level depending on the location) and thus was selected and used as the low-income household threshold in this study.

Table 1. Summary of low-income definitions.

| Source | Threshold Name | Geographical Resolution | HH Properties | 2017 HH Low-income Threshold for 4 Person HHs in NYC |
|---|---|---|---|---|
| US Census Bureau (2017a) | Census Bureau Poverty Threshold | Nation | Household Size & Composition | $25,094 |
| Department of Health and Human Services (2017) | HHS Poverty Guidelines | Nation | Household Size | $24,600 |
| US Census Bureau (2017b) | Census Bureau SPM | Metropolitan Area | Household Size & Housing Tenure Status | $31,672 (2 adult 2 child) $38,737 (4 adult 0 child) |
| Department of Labor (2017) | LLSIL Guidelines | Selected Metropolitan Areas | Household Size | $31,852 |
| US Department of Housing and Urban Development (2017) | HUD Income Limits | County/MSA | Household Size | Very low-income: $47,700 Low income: $76,300 |

The HUD developed low-income/very low-income thresholds to determine the eligibility for assisted housing programs that include Public Housing. The income limits are set based on HUD estimates of median family income (MFI) at each fiscal year. Very low-income family is defined as those with incomes that do not exceed 50 percent of the median family income for the areas while 80 percent is chosen as the threshold for low-income families. Comparing the two definitions, it was found that the low-income threshold (80% of MFI) classifies nearly 50% of NYS HHs as low-income HHs which provides less value to low-income studies. Therefore, in this study, the HUD very low-income family definition was used to define the low-income HHs. About 30% of NYS HHs were classified as low-income HHs based on this threshold. All subsequent mention of low-income HHs was defined using the HUD very low-income family threshold.

**METHODOLOGY**

The objective of this study was to explore how the mobility differences between low-income HHs and their counterparts vary among distinct socio-demographic groups. To achieve this goal, socio-demographic groups need to be defined first. Socio-demographic groups in literature are generally self-defined based on one or two attributes (e.g., elderly household vs non-elderly HHs, elderly HHs in the urban areas vs non-elderly HHs in urban areas). If more attributes need to be considered, the full combination of these attributes results in a long list of household categories, which is not preferred. To this end, this study adopted a clustering approach to automatically identify the distinct socio-demographic groups based on attributes that were found to have impacts on low-income household travel behavior. These attributes were identified based on an explorative analysis of the 2017 NHTS statistics in NYS. Table 2 summarizes the nine socio-demographic attributes as well as the associated variable types and descriptive statistics among the entire population in NYS. Among the nine variables, household size and household vehicle ownership are numerical variables. The remaining seven variables are divided into different categories as presented in the descriptive statistics column. For example, the household location is classified into three categories: NYC, other NYS urban areas other than NYC and rural areas. The share of the households among each group is presented as well.

**Table 2. Summary statistics of low-income households socio-demographic variables.**

| Variable | Type | Descriptive Statistics |
|---|---|---|
| Household size | Numerical | Median: 2 |
| Household vehicle ownership | Numerical | Median: 2 |
| Household location | Categorical | 8.41% NYC, 70.28% other urban, 21.31% rural |
| Elderly status | Categorical | 40.66% elderly household, 49.34% non-elderly household |
| Household race | Categorical | 89.86% white, 10.14% non-white |
| Employment status | Categorical | 64.16% working household, 35.84% non-working household |
| Education status | Categorical | 82.42% higher educated household, 17.58% lower educated household |
| Gender distribution | Categorical | 31.63% #males < #females, 44.94% #males = #females, 23.43% #males > #females |
| Vehicle/driver distribution | Categorical | 11.49% #vehicles < #drivers, 68.04% #vehicles = #drivers, 20.47% #vehicles > #drivers |

Note: Urban is defined as metropolitan areas in NYS
An elderly household is defined as at least one household member is 65 or older
Working household is defined as at least one household member is employed
Higher education is defined as a college or higher degree and vice versa

A K-prototype algorithm (Huang 1998) was employed in this study to categorize the sample data based on the above socio-demographic attributes. The K-prototype algorithm is an improvement of the K-Means and K-Mode clustering algorithm to handle clustering with mixed data types such as a mixture of categorical and numerical variables. The algorithm was implemented using Python kmodes library (De Vos 2022).

**RESULTS AND DISCUSSION**

**Socio-Demographic Clusters**

Figure 1 displays the number of clusters and the average dispersion. The average dispersion is based on a cost function that considers the sum of distances of all points to their respective cluster centroids. To determine the optimal number of clusters, the elbow method was used. Based on the figure, 5 was selected as the optimal cluster number because the marginal benefit in average dispersion (decrease in average dispersion) is not significant once the number of clusters exceeds 5.

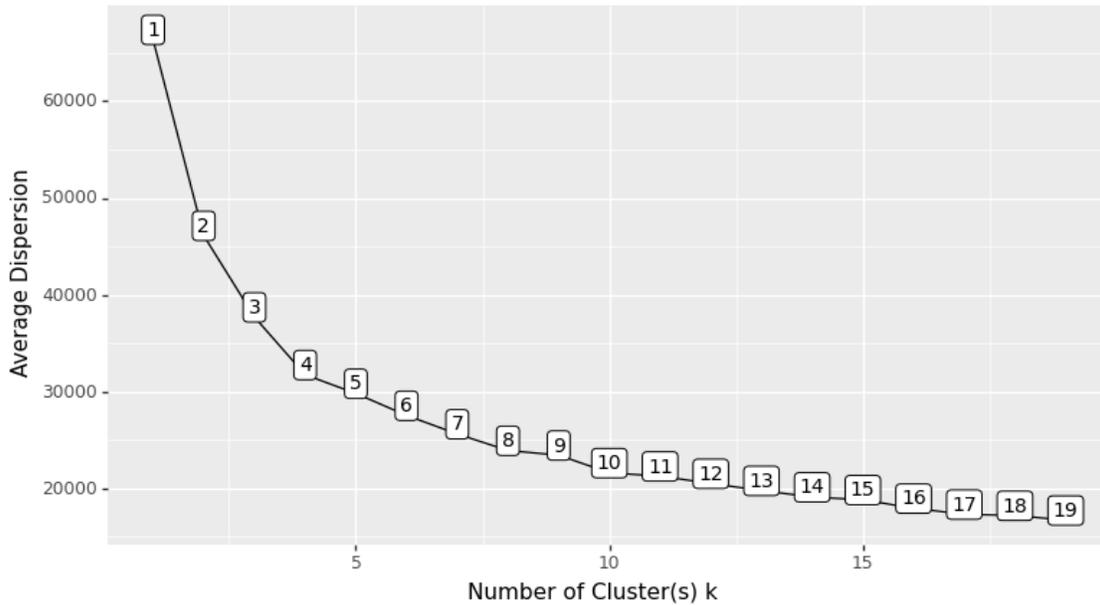

**Figure 1. Number of clusters by average dispersion.**

Table 3 summarizes the socio-demographic variable statistics within each cluster. For the two numeric variables, the statistics presented is the median value. For example, the median household size for Cluster 1 is 2 people. The proportion of each category is presented for the categorical variables. Take Cluster 1 household location as an example, 69.6% of households in this cluster lived in other urban areas while 28.6% lived in rural NYS areas.

**Table 3. Summary statistics for variables by cluster.**

| Variable | Cluster 1 | Cluster 2 | Cluster 3 | Cluster 4 | Cluster 5 | All |
|---|---|---|---|---|---|---|
| HH size (median) | 2 | 2 | 4 | 1 | 1 | 2 |
| HH vehicle ownership (median) | 3 | 2 | 2 | 1 | 1 | 2 |
| HH location (%other urban) | 69.6 | 73.1 | 74.2 | 64.9 | 67.4 | 70.3 |
| HH location (%NYC) | 1.8 | 5.1 | 8 | 10.5 | **16.3** | 8.4 |
| HH location (%rural) | 28.6 | 21.8 | 17.8 | 24.6 | 16.2 | 21.3 |
| Elderly status (%elderly HH) | 36.3 | 60.8 | 10.8 | **75.6** | 14.5 | 40.7 |
| Elderly status (%non-elderly HH) | 63.7 | 39.2 | **89.3** | 24.4 | 85.5 | 59.3 |
| HH race (%white) | **95.1** | 94.2 | 86.7 | 86.9 | 85.2 | 89.9 |
| HH race (%non-white) | 4.9 | 5.8 | 13.3 | 13.1 | 14.8 | 10.1 |
| Employment status (%working HH) | 76.7 | 58.6 | 95 | 8.1 | 79.9 | 64.2 |
| Employment status (%non-working HH) | 23.3 | 41.4 | 5 | 91.9 | 20.1 | 35.8 |
| Education status (%higher educated HH) | 88.6 | 87.2 | **91.8** | 46.3 | 91.3 | 82.4 |
| Education status (%lower educated HH) | 11.4 | 12.8 | 8.2 | 53.7 | 8.7 | 17.6 |
| Gender distribution (%#males < #females) | 12.6 | 6.3 | 41 | 72.2 | 40.5 | 31.6 |
| Gender distribution (%#males = #females) | 68.4 | **86.4** | 27.5 | 7.4 | 15.1 | 44.9 |
| Gender distribution (%#males > #females) | 19 | 7.3 | 31.5 | 20.4 | **44.4** | 23.5 |
| Vehicle/driver distribution (%#vehicles < #drivers) | 0 | 10.9 | 18 | 8.6 | 16.5 | 11.5 |
| Vehicle/driver distribution (%#vehicles = #drivers) | 2.4 | 81.7 | 73.6 | **85.8** | 74.4 | 68 |
| Vehicle/driver distribution (%#vehicles > #drivers) | **97.6** | 7.4 | 8.4 | 5.6 | 9.1 | 20.5 |

Examining the statistics presented in Table 3, some representative socio-demographic statistics in each cluster were identified and highlighted. Each cluster is given a name based on the representative statistics as presented in Table 4. The first cluster is dominated by HHs that had more vehicles than drivers. Over 14% of the total NYS HHs are part of this cluster. The majority (77%) of HHs in Cluster 1 were not low-income HHs. Cluster 2 includes HHs with equal male and female residents. The median household size is 2. Cluster 3 is dominated by higher educated non-elderly HHs. 95% of these HHs were working HHs. As expected, Cluster 3 has a much lower proportion of low-income HHs. The majority of Cluster 4 are elderly HHs with more females than males and have equal vehicles and drivers. These HHs had lower education status compared to other clusters. The share of the low-income HHs in Cluster 4 was the highest—over 70% of them had an income below the low-income threshold. Lastly, Cluster 5 is NYC HHs with more male than female residents—most of these were 1-person HHs.

**Table 4. Names and properties of clusters.**

| Cluster | Cluster name | Share (sample size) | Low-income vs not low-income share |
|---|---|---|---|
| 1 | White household with more vehicles than drivers | 14.2% (2,339) | Not low-income dominated (77.1% not low-income) |
| 2 | Household with equal male and female residents | 29.8% (4,893) | Not low-income dominated (88.6% not low-income) |
| 3 | Higher educated non-elderly household | 18.6% (3,063) | Not low-income dominated (85.1% not low-income) |
| 4 | Elderly household with equal vehicles and drivers | 16.4% (2,691) | Low-income dominated (71.7% low-income) |
| 5 | NYC household with more male than female residents | 21.1% (3,466) | Not low-income dominated (81.0% not low-income) |

The mobility difference between low-income HHs and their not low-income neighbors, including average daily person trips, person miles traveled, average trip length, and trip duration, were examined among each demographic group (i.e., cluster). Note that the differences were calculated as the statistics in not low-income HHs minus those in low-income HHs. Figures 2 through 5 summarize all the mobility differences.

Figure 2 shows the average daily person trip differences in each group. Overall, the residents from low-income HHs made fewer personal trips compared to their not low-income counterparts. The differences are most obvious within group 1 (white household with more vehicles than drivers) and group 4 (elderly household with equal vehicles and drivers). However, while the daily trip gaps are significant for the entire population, they are not statistically significant at any individual group level.

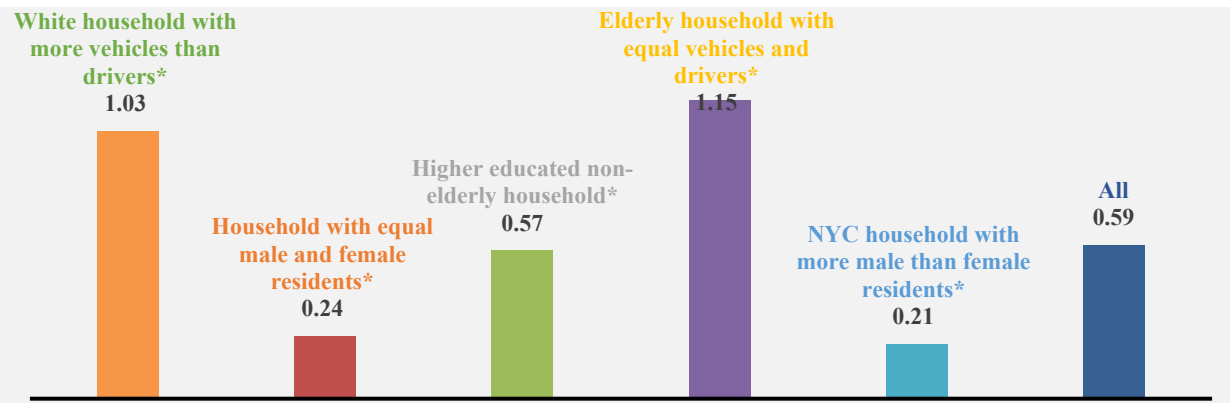

**Figure 2. Average daily person trip differences (Not low-income HHs − Low-income HHs).**
*Note: * not statistically significant at 5% confidence level.*

Figure 3 shows the person miles traveled (PMT) differences in miles between the two income groups cross each demographic group. On average, the daily PMT made by residents from low-income HHs is 12.4 miles less than that of residents from not low-income HHs. Among all demographic groups, the white household with more vehicles than drivers shows the largest differences. In particular, Low-income household members generally traveled 21.5 miles less per day than their not low-income counterparts. No obvious difference can be found in the NYC household with more male than female resident group.

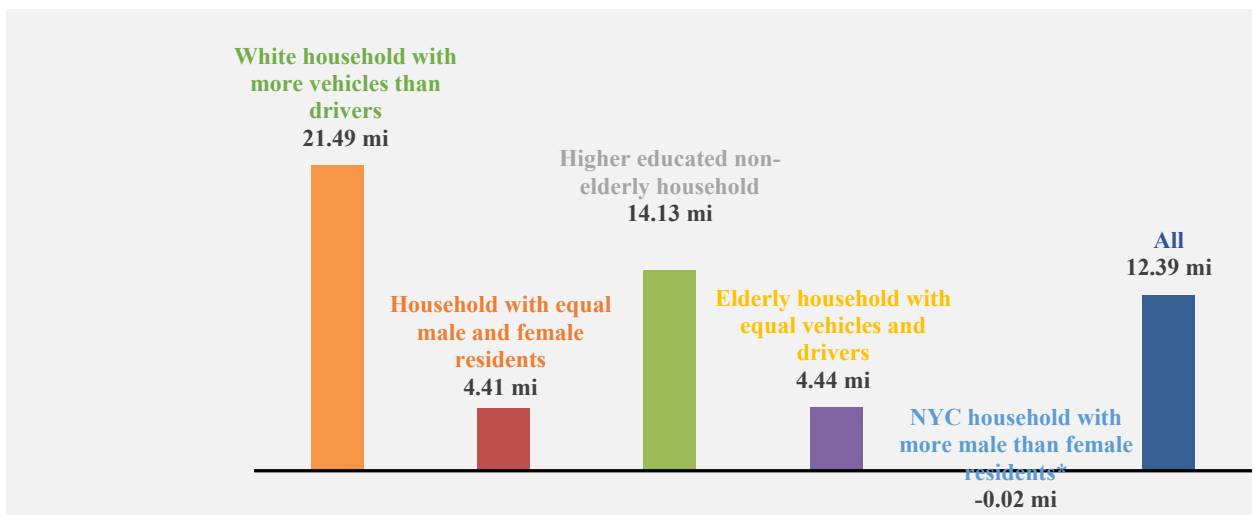

**Figure 3. Average daily PMT differences in miles (Not low-income HHs − Low-income HHs).**
*Note: * not statistically significant at 5% confidence level.*

Figure 4 shows the trip length differences per trip between the two income groups. Note that trips by air were excluded in this trip length calculation. Similar to PMT, the trip length gaps within white household with more vehicles than drivers and higher educated non-elderly HH are most evident. The trip length gaps in the other three socio-demographic groups are not statistically significant. Overall, the average trip length made by a person from a low-income household is 2.7 miles shorter than from not low-income HHs in NYS.

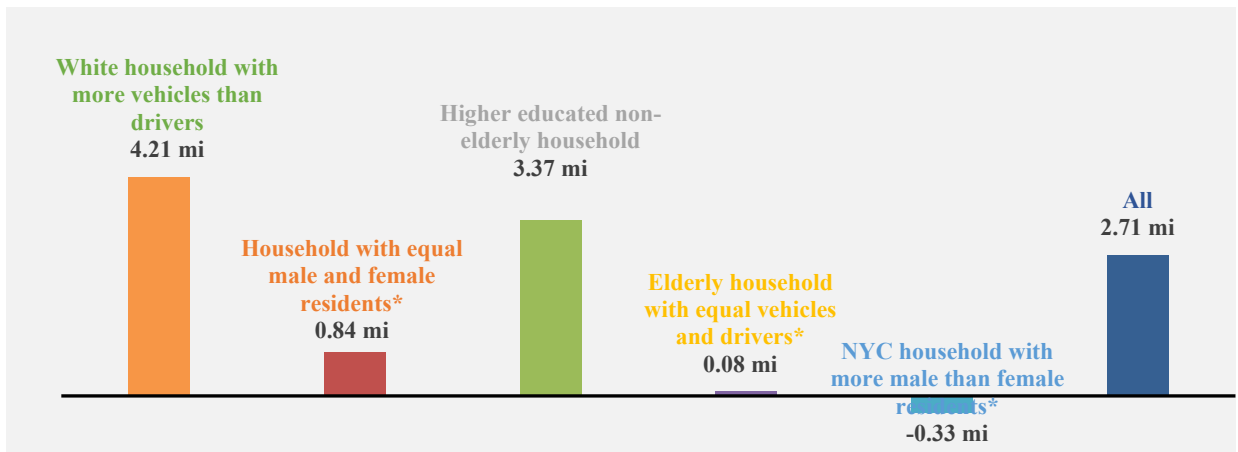

**Figure 4. Trip length differences in miles (Not low-income HHs − Low-income HHs).**
*Note: * not statistically significant at 5% confidence level.*

Figure 5 presents the average trip duration differences within each group. Air trips were excluded from this trip duration calculation. In contrast to the average daily person trip length, the average trip duration for residents from low-income HHs was longer than that of residents from not low-income HHs. Possible reasons could be that the low-income HHs either suffered more from traffic congestion or took less time-efficient modes such as public transit or walking instead of driving. However, the difference in each group varies. The disparity was most obvious among elderly household with equal vehicles and drivers. On the other hand, for those from white household with more vehicles than drivers, the average trip duration was higher for residents from not low-income HHs compared to that of low-income HHs.

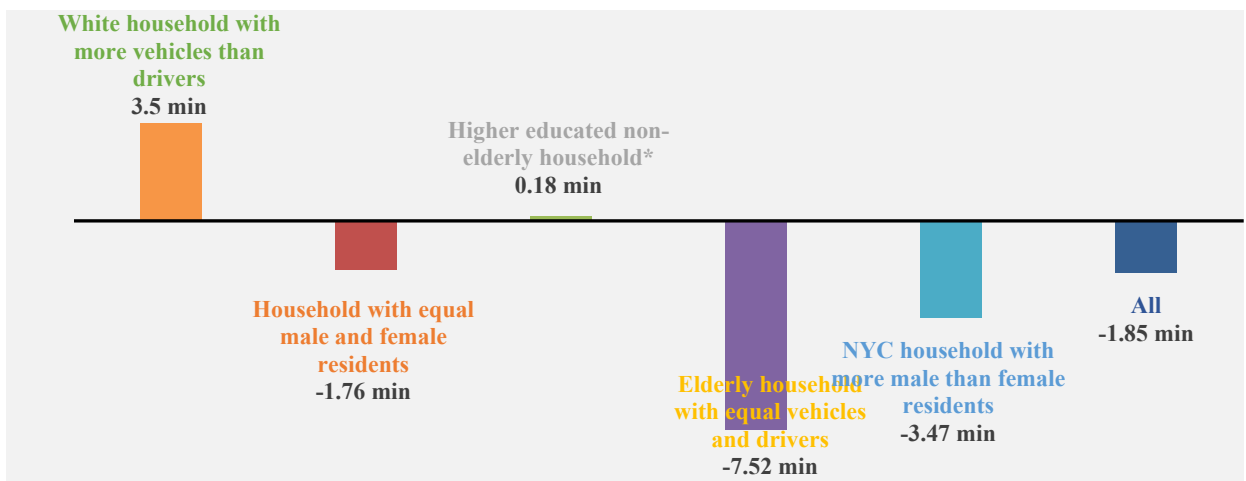

**Figure 5. Trip duration differences in minutes (Not low-income HHs − Low-income HHs).**
*Note: * not statistically significant at 5% confidence level.*

**CONCLUSION**

Using the 2017 National Household Travel Survey, this study investigated the mobility gaps between low-income and not low-income households (HHs) in New York State (NYS). Different from previous studies which use one single low-income threshold for the entire study area, the low-income threshold of 50% of the median income, published by Housing and Urban Development (HUD), was adopted in this study. HUD establishes the threshold in county/metropolitan areas which captures the cost of living in different regions and is considered as a more reasonable measurement than those established at the nation level. Furthermore, to better understand the mobility gaps among different socio-demographic groups, a K-prototype clustering approach was adopted to categorize the population based on various attributes, such as household size, vehicle ownership, gender, employment status, and education. Then, the mobility differences among the entire population in NYS as well as within each population group were examined. Results from the analysis confirmed the findings from other studies that individuals from low-income HHs generally made fewer trips and shorter trip distances compared to their not low-income counterparts. The trip length differences as well as person miles traveled between low-income HHs and non-low-income HHs were most obvious among the white HHs with more vehicles than drivers. Although the residents from low-income HHs made shorter trips on average, they experienced longer travel time than those from not low-income HHs.

In future studies, the authors aim to evaluate other travel behavior gaps (e.g., transportation mode and trip purpose) between low-income and not low-income HHs. In addition, the authors plan to apply the methodology used in this study to other geographical regions to evaluate whether the mobility gaps within distinct socio-demographic groups vary from region to region.

**ACKNOWLEDGMENTS**

This manuscript has been authored by UT-Battelle, LLC, under contract DE-AC05-00OR22725 with the US Department of Energy (DOE). The US government retains and the publisher, by accepting the article for publication, acknowledges that the US government retains a nonexclusive, paid-up, irrevocable, worldwide license to publish or reproduce the published form of this manuscript, or allow others to do so, for US government purposes. DOE will provide public access to these results of federally sponsored research in accordance with the DOE Public Access Plan (http://energy.gov/downloads/doe-public-access-plan).

The authors would like to acknowledge the support of the New York State Department of Transportation (NYSDOT). The opinions, findings, and conclusions in this
paper are those of the authors and not necessarily those of the NYSDOT.